\documentclass[11pt]{article}

\usepackage{mathrsfs}
\usepackage{amssymb}
\usepackage{graphicx}
\usepackage{chet}

\newcommand{\lsp}{\hspace{1pt}}

\title{Symplectic critical models in $\titlemath{6+\epsilon}$ dimensions}

\author{Andreas Stergiou\emails{andreas.stergiou@yale.edu}}

\affiliation{Department of Physics, Yale University, New Haven, CT 06520,
USA}

\abstract{We consider nontrivial critical models in $d=6+\epsilon$
spacetime dimensions with anticommuting scalars transforming under the
symplectic group $\text{Sp}(N)$. These models are nonunitary, but the
couplings are real and all operator dimensions are positive. At large $N$
we can take $\epsilon\to1$ consistently with the loop expansion and thus
provide evidence that these theories may be used to define critical models
in $d=7$. The relation of these theories to critical $\text{Sp}(N)$
theories, defined similarly to the well-known critical $\text{O}(N)$
theories, is examined, and some similarities are pointed out.}

\date{August 2015}

\begin{document}

\maketitle

\newsec{Introduction}
Conformal field theories (CFTs) in $d=2$ spacetime dimensions are abundant
and their properties have been studied extensively. As we consider higher
spacetime dimensions it becomes harder to find nontrivial CFTs, and if we
require supersymmetry it becomes impossible beyond
$d=6$~\cite{Nahm:1977tg}. In this short paper we will give up on some basic
properties of CFTs in $d\le6$ in order to provide evidence for the
existence of (perhaps unconventional) interacting CFTs in $d=7$.

Our considerations are inspired by recent work of Fei \emph{et.\ al.}, who
analyzed $\text{O}(N)$ and $\text{Sp}(N)$ models in $d=6-\epsilon$ in great
detail~\cite{Fei:2014yja, Fei:2015kta}. For the $\text{Sp}(N)$ case they
worked with the Lagrangian
\eqn{\mathscr{L}=\tfrac12\lsp\Omega_{ij}\lsp\partial^\mu\chi^i\lsp
\partial_\mu\chi^j+\tfrac12\lsp\partial^\mu\sigma\lsp\partial_\mu\sigma
+\tfrac12\lsp g\lsp\Omega_{ij}\lsp\chi^i\chi^j\sigma
+\tfrac16\lsp h\lsp\sigma^3\,,}[Lag]
where $\Omega_{ij}$ is the invariant symplectic matrix. The scalar fields
$\chi$ are anticommuting, and so the theory is not unitary, while the
scalar field $\sigma$ is commuting. We point out that in $d=6+\epsilon$
these $\text{Sp}(N)$ models have UV fixed points at real values of the
couplings and with positive operator dimensions. The corresponding theories
have a potential that is unbounded from below, but within perturbation
theory the vacuum configuration $\chi=\sigma=0$ is stable. This is similar
to the situation in~\cite{Fei:2014yja}.

A critical theory with $\text{O}(N)$ symmetry and commuting scalars that
can formally be defined in any $d$ also exists, and its central charge has
been computed analytically in $d$ at leading order in
$1/N$~\cite{Petkou:1994ad, Petkou:1995vu}. If we send $N\to-N$ in the
answer, then this gives the central charge of the corresponding critical
theory with $\text{Sp}(N)$ symmetry and anticommuting scalars.  For $d=7$
the value of the central charge indicates the presence of an interacting
fixed point.

It is not clear if the corresponding CFT is related to the theory defined
at the critical point of \Lag we discuss in this paper.  In the context of
the $1/N$ expansion one can instead study the (Euclidean) theory
\eqn{\mathscr{L}_{\text{sym}}=\tfrac12\lsp\Omega_{ij}\lsp
\partial^\mu\chi^i\lsp\partial_\mu\chi^j
+\tfrac14\lambda\lsp(\Omega_{ij}\lsp\chi^i\chi^j)^2\,.}[Lsym]
(For details of the corresponding $\text{O}(N)$ theory at large $N$ the
reader is refererd to~\cite{Moshe:2003xn}). Typically one does this in
$d=4-\epsilon$ dimensions, but one can also use $d=4+\epsilon$. In the
latter case we can take $\epsilon\to3$ at large $N$ and we have a UV fixed
point as well. This fixed point is nonunitary due to the anticommuting
scalars, and there is a violation of the unitarity bound for $\chi$.
Nevertheless it has real coupling $\lambda$ and positive operator
dimensions at large $N$, so it is very similar in nature to the nontrivial
fixed point of \Lag.

The paper is organized as follows. In the next section we analyze the fixed
points of \Lag borrowing heavily on results of~\cite{Fei:2014yja,
Fei:2014xta, Fei:2015kta}.  In section~\ref{CC} we consider the central
charge of the critical $\text{Sp}(N)$ models in $d>6$, and we speculate on
the relation of the corresponding CFT to the theory at the critical point
of \Lag.  We make some comments on the sphere free energy and the
$F$-theorem~\cite{Jafferis:2010un, Jafferis:2011zi, Klebanov:2011gs} in
section~\ref{sfe}, and we conclude in section~\ref{conc}.

\newsec{Fixed points}
In $d=6+\epsilon$ and at one loop the beta functions for $g$ and $h$
are~\cite{deAlcantaraBonfim:1981sy, Fei:2014yja, Fei:2015kta,
Grinstein:2015ina}
\twoseqn{\beta_g&=\frac{\epsilon}{2}\lsp g-\frac{1}{12}
\frac{1}{64\lsp\pi^3}\lsp g\big((N+8)g^2+12\lsp gh-h^2\big)\,,
}[betagFull]{\beta_h&=\frac{\epsilon}{2}\lsp h
+\frac14\frac{1}{64\lsp\pi^3}(4\lsp N g^3-N g^2 h-3\lsp h^3)\,.
}[betahFull][betasFull]
In the large-$N$ limit we have
\twoseqn{\beta_{g,\lsp N\gg1}&=\frac{\epsilon}{2}\lsp g-\frac{1}{12}
\frac{1}{64\lsp\pi^3}\lsp N g^3\,,}[betag]
{\beta_{h,\lsp N\gg1}&=\frac{\epsilon}{2}\lsp h
+\frac14\frac{1}{64\lsp\pi^3}Ng^2(4\lsp g-h)\,,}[betah][betas]
and it is easy to find nontrivial fixed points.  For these fixed points
higher loop corrections in \betag and \betah can be neglected consistently
at large $N$, for, due to the interactions in \Lag, each higher order in
perturbation theory generates contributions to the beta function that are
at most linear in $N$. One nontrivial fixed point at large $N$ is at
\eqn{g_\ast=8\sqrt{6}\lsp\pi^{3/2}\sqrt{\frac{\epsilon}{N}}\,,\qquad
h_\ast=6\lsp g_\ast\,,}[fp]
and there is also an equivalent fixed point at $(-g_\ast,-h_\ast)$. Since
$\epsilon,N>0$ these fixed points occur for real values of the couplings.
Corrections in powers of $1/N$ that give solutions to $\beta_g=\beta_h=0$
can also be computed and give~\cite{Fei:2014yja}
\eqn{g_\ast=8\sqrt{6}\lsp\pi^{3/2}\sqrt{\frac{\epsilon}{N}}
\left(1-\frac{22}{N}+\cdots\right)\,,\qquad h_\ast=
48\sqrt{6}\lsp\pi^{3/2}\sqrt{\frac{\epsilon}{N}}
\left(1-\frac{162}{N}+\cdots\right)\,,}[fpfull]
while higher loop corrections have been considered
in~\cite{deAlcantaraBonfim:1981sy, Grinstein:2015ina, Fei:2014xta}. Note
that the next-to-leading-order result in $1/N$ in \fpfull is sensitive to
higher-loop corrections in the limit $\epsilon\to1$~\cite{Fei:2014xta}.
These corrections are $\mathcal{O}(\epsilon^{3/2})$ in \fpfull.

The eigenvalues of the stability matrix at the fixed point \fp are negative
(both equal to $-\epsilon$), so these are UV fixed points. The trivial
fixed point is an IR fixed point. A ``UV completion'' of the nontrivial
fixed points, so that they appear as IR fixed points of another theory
needs to be considered, although at this point the existence of such a ``UV
completion'' is unclear. Following the example of~\cite{Fei:2014yja} one
may speculate that this may be found starting from a theory in
$d=8-\epsilon$.

The anomalous dimensions of the fields at one loop are given
by~\cite{deAlcantaraBonfim:1981sy, Fei:2014yja, Fei:2015kta,
Grinstein:2015ina}
\twoseqn{\gamma_\chi&=\frac16\frac{1}{64\lsp\pi^3} g^2\,,}[gammaphi]
{\gamma_\sigma&=-\frac{1}{12}\frac{1}{64\lsp\pi^3}\lsp (N
g^2-h^2)\,.}[gammasigma][gammas]
At the fixed point \fpfull we have
\eqn{\gamma_{\chi\lsp\ast}=\frac{\epsilon}{N}\left(1-\frac{44}{N}+
\cdots\right)\,,\qquad
\gamma_{\sigma\lsp\ast}=-\frac{\epsilon}{2}+\frac{40\lsp\epsilon}{N}
\left(1-\frac{170}{N}+\cdots\right)\,.}[gammafp]
Neglecting higher-loop effects (of $\mathcal{O}(\epsilon^2)$ in \gammafp)
and using the result up to order $1/N$ we see, in the limit $\epsilon\to1$,
that the anomalous dimension of $\sigma$ is negative if $N\ge80$. In that
case the unitarity bound is violated for $\sigma$.  Nevertheless, the
violation is mild and the dimension of $\sigma$ is positive. If three-loop
effects are taken into account then~\cite{Fei:2014xta}
\eqna{\gamma_{\chi\lsp\ast}&=\frac{\epsilon}{N}\left(1
+\frac{11}{12}\lsp\epsilon
-\frac{13}{144}\lsp\epsilon^2+\mathcal{O}(\epsilon^3)\right)
+\mathcal{O}\left(\frac{1}{N^2}\right)\,,\\
\gamma_{\sigma\lsp\ast}&=-\frac{\epsilon}{2}+\frac{40\lsp\epsilon}{N}
\left(1+\frac{13}{15}\lsp\epsilon-\frac{11}{180}\lsp\epsilon^2
+\mathcal{O}(\epsilon^3)\right)+\mathcal{O}\left(\frac{1}{N^2}\right)\,.
}[gammafpII]
With the result leading in $1/N$ and neglecting the
$\mathcal{O}(\epsilon^4)$ contributions we see that in the limit
$\epsilon\to1$ the anomalous dimension of $\chi$ is positive for all
positive $N$, while that of $\sigma$ is positive for positive
$N\le\frac{1300}{9}\approx 144.4$.

\begin{figure}[ht]
  \centering
  \includegraphics{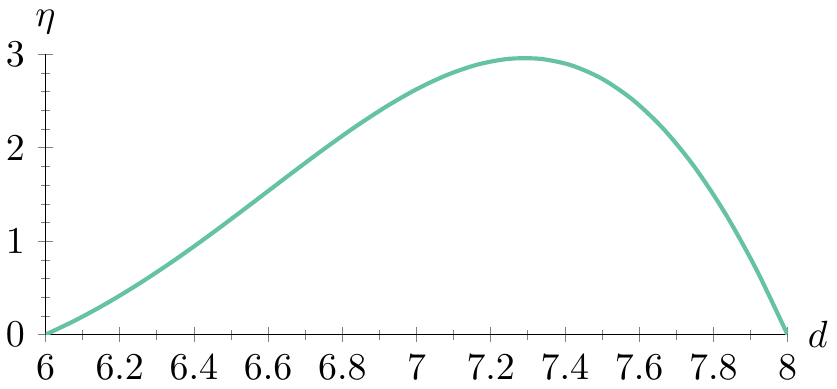}
  \caption{Plot of the function $\eta(d)$ defined in \etadef for $6\le
  d\le8$.} \label{fig:eta}
\end{figure}

The result \gammafpII can be improved for results analytic in $d$ for the
anomalous dimensions of $\chi$ and $\sigma$ at large $N$ also
exist~\cite{Vasiliev:1981yc, Vasiliev:1981dg, Vasiliev:1982dc,
Fei:2014yja}. At leading order in $1/N$ they are
\eqn{\gamma_{\chi\lsp\ast}=\frac{1}{N}\lsp\eta\,,\qquad
\gamma_{\sigma\lsp\ast}=-\frac{d-6}{2}+\frac{1}{N}
\frac{4\lsp(d-1)(d-2)}{d-4}\lsp\eta\,,}[Deltas]
where
\eqn{\eta=-\frac{2^{d-3}(d-4)\lsp\Gamma(\frac{d-1}{2})\sin\frac{\pi
d}{2}}{\pi^{3/2}\lsp\Gamma(\frac12\lsp d+1)}\,.}[etadef]
The function $\eta=\eta(d)$ is plotted in Fig.~\ref{fig:eta} in the region
$6\le d\le8$.
As we see $\eta$ is zero at $d=6,8$. This suggests that $\chi$ becomes free
in $d=8$, while $\sigma$ has $\Delta_\sigma=2$, well below the unitarity
bound. The positivity of $\eta$ in the region $6<d<8$ indicates that if we
trust the $1/N$ expansion at $N$ not too large then the dimension of
$\sigma$ may not violate the unitarity bound for $6<d<8$. For $d=7$, for
example, the unitarity bound for $\sigma$ is violated if
$N>\frac{8192}{7\lsp\pi^2}\approx 118.6$.\foot{I thank Simone Giombi and
Igor Klebanov for suggesting the analysis of $\gamma_\chi$ and
$\gamma_\sigma$ presented here.}

One can also consider operator mixing between $\Omega_{ij}\chi^i\chi^j$ and
$\sigma^2$. In the large-$N$ limit the results can again be borrowed
from~\cite{Fei:2015kta}. From the equation of motion of $\sigma$ it is
clear that there is one linear combination of $\Omega_{ij}\chi^i\chi^j$ and
$\sigma^2$ that is a descendant of $\sigma$.  Its scaling dimension is
$\Delta_\sigma+2$. The other independent combination is a primary with
scaling dimension $\Delta=d-2-\frac{100\lsp\epsilon}{N}$.

We note here that for $N=2$ there are fixed points of \betagFull and
\betahFull with a symmetry enhancement from $\text{Sp}(2)$ to the
orthosymplectic supergroup $\text{OSp}(1|2)$~\cite{Fei:2015kta}.
Considering $\epsilon\to1$ this points to the existence of
$\text{OSp}(1|2)$ symmetric CFTs in $d=7$, although here we are no longer
in the large-$N$ limit.

Our discussion provides evidence for the existence of nontrivial CFTs in
$d=7$. In the large-$N$ limit the $\epsilon\to 1$ limit of our
$d=6+\epsilon$ results can be taken consistently with neglecting
higher-loop effects. While the CFTs for which we find evidence are not
unitary, the violation of unitarity is not due to imaginary
couplings.\foot{This would be the case if we had $\text{O}(N)$ symmetry
with commuting scalars $\phi^i$ in \Lag in $d=6+\epsilon$.} Distinctively,
all operator dimensions are positive, although that of $\sigma$ violates
the unitarity bound at large $N$.  This violation is mild, but it still
shows that not all states in the theory have positive norm.  As we see from
\Deltas $\Delta_\sigma\to2$ as $N\to\infty$ in all $d$.

\newsec{Central charge}[CC]
The two-point function of the stress-energy tensor can be defined as
\eqn{\langle T_{\mu\nu}(x)\lsp T_{\rho\sigma}(0)\rangle=
C_T\lsp\frac{I_{\mu\nu\hspace{-0.5pt}\rho\sigma}(x)}{x^{2d}}\,,}[]
with
\eqn{I_{\mu\nu\hspace{-0.5pt}\rho\sigma}(x)=\frac12\lsp\big(I_{\mu\rho}(x)
I_{\nu\sigma}(x)+I_{\mu\sigma}(x)I_{\nu\rho}(x)\big)
-\frac{1}{d}\lsp\delta_{\mu\nu}\lsp\delta_{\rho\sigma}\,,
\qquad I_{\mu\nu}(x)=\delta_{\mu\nu}-2\lsp\frac{x_\mu x_\nu}{x^2}\,.}[]
The central charge of the critical $\text{O}(N)$ theory with commuting
scalars is given by~\cite{Petkou:1994ad, Petkou:1995vu}
\eqn{C_T=\frac{d\lsp\Gamma^2(\frac12\lsp d)}{4(d-1)\lsp\pi^d}
\left(N+\left(\frac{4\lsp \mathcal{C}(\frac12\lsp d)}{d+2}
+2\frac{d^{\hspace{0.5pt}2}+6\lsp d-8}{d(d-2)(d+2)}\right)\eta
+\mathcal{O}\left(\frac{1}{N}\right)\right)\,,}[cc]
where $\eta$ is given in \etadef and
\eqn{\mathcal{C}(x)=\psi(3-x)+\psi(2\lsp x-1)-\psi(x)-\psi(1)\,,\qquad
\psi(x)=\frac{\Gamma'(x)}{\Gamma(x)}\,.}[]
Setting $d=6-\epsilon$, expanding in $\epsilon$ and taking $\epsilon\to0$
we find
\eqn{C_{T,\lsp d=6}=\frac{6}{5\lsp\pi^6}(N+1)\,,}[CTsix]
and so in $d=6$ we get a result for the central charge consistent with
$N+1$ free scalars. This was discussed in~\cite{Fei:2014yja} as a nice
check of their proposal for the UV completion of the $d=5$ critical
$\text{O}(N)$ theory.

If $N$ scalars are anticommuting then $N\to-N$ in \CTsix. In the remainder
of this section we send $N\to-N$ and apply \cc in $d>6$. This corresponds
to the critical $\text{Sp}(N)$ symmetric theory with anticommuting scalars.

From \cc we can see that, if we neglect $1/N$ corrections, then $C_T$ is
negative for all integer $N\ge2$ in the limit $\epsilon\to1$, for then
\eqn{C_{T,\lsp d=7}=-\frac{525}{512\lsp\pi^6}\lsp\left(N-\frac{516\lsp
608}{33\lsp075\lsp\pi^2}\right)\,.}[CTres]
This result indicates the presence of an interacting fixed point. It is not
clear if this fixed point is related to the $\epsilon\to1$ limit of the
$d=6+\epsilon$ fixed point of \Lag discussed here. Nevertheless, if we
identify it with the fixed point of \Lsym in the context of the $1/N$
expansion, then we observe some similarities: both fixed points are
nonunitary with violations of the unitarity bounds for operator dimensions,
and they both have real couplings and positive operator dimensions.

Let us also consider $d=8$ where we get
\eqn{C_{T,\lsp d=8}=-\frac{72}{7\lsp\pi^8}\lsp(N+4)\,,}[CTeight]
indicating a free theory there. This theory in $d=8-\epsilon$ may have an
IR fixed point. This may coincide with the UV fixed point of \Lag in $d=7$,
something that would provide its UV completion.

For the critical $\text{O}(N)$ models we see a shift of $N$ by $1$ in
\CTsix and by $-4$ in \CTeight. Although $\eta(d)$ is zero for even $d$,
there is an obvious pole at $d=2$ in \cc, while at even $d\ge6$
$\mathcal{C}(\frac12\lsp d)$ has a pole due to $\psi(3-\frac12\lsp d)$.
Consequently, $N$ is shifted in \cc in $d=2$ and in even $d\ge6$.  The
analytic expression that gives the shift of $N$ in even $d\ge2$ is\foot{I
thank David Poland for collaboration on this calculation.}
\eqn{N_{\text{shift}}=-\frac{(d-4)\lsp\Gamma(d-1)} {\Gamma(\frac12
d)\Gamma(\frac12 d+2)}\cos\frac{\pi d}{2}\,.}[]
This is an integer for any even $d\ge2$.

\newsec{Sphere free energy and the \texorpdfstring{$F$}{F}-theorem}[sfe]
At the interacting fixed point in $d=6+\epsilon$ we can compute the sphere
free energy using the results of~\cite{Giombi:2014xxa}. If $Z_{S^d}$ is the
partition function on the $d$-dimensional sphere, then for
\eqn{F=-\log Z_{S^d}\,,\qquad\tilde{F}=-F\sin\frac{\pi d}{2}\,,}[]
we have, for the interacting theory,
\eqn{F=F_{\text{free}}
+\frac{1}{8640}\frac{1}{64\lsp\pi^3}
(h_\ast^2-3\lsp N g_\ast^2)+\mathcal{O}(\epsilon^2)\,,}[]
and
\eqn{\tilde{F}=\tilde{F}_{\text{free}}
+\frac{\pi}{17\lsp280}\frac{h_\ast^2-3\lsp N g_\ast^2}
{64\lsp\pi^3}\lsp\epsilon+\mathcal{O}(\epsilon^3)\,,}[]
where $F_{\text{free}}=(1-N)F_s$ with $F_s$ the value of $F$ for a free
conformal scalar. In our examples in $d=6+\epsilon$ the IR theory is free
and the UV interacting. As a result, in the large-$N$ limit we get
\eqn{F_{\text{UV}}<F_{\text{IR}}\,,\qquad
\tilde{F}_{\text{UV}}<\tilde{F}_{\text{IR}}\,,}[Fthm]
for a flow between these two theories.  One can also verify that \Fthm
holds for any $N\ge2$~\cite{Fei:2015kta}.  Note that results analytic in
$d$ also exist here~\cite{Giombi:2014xxa}.

As we see the $F$-theorem is violated both for $F$ and for $\tilde{F}$ in
$d=6+\epsilon$.  This does not raise concerns for the validity of the
$F$-theorem since the theory we are considering is not unitary. A closely
related example, where the $F$-theorem for $F$ is violated but, contrary to
our case, that for $\tilde{F}$ holds despite the violation of unitarity,
was encountered in~\cite{Fei:2015kta}.

\newsec{Conclusion}[conc]
In this short paper we provided evidence for the existence of nonunitary UV
fixed points in $d=6+\epsilon$ dimensions, and suggested that these fixed
points survive in $d=7$. The important distinguishing feature of the fixed
points we propose is that the critical values of the couplings are real and
the operator dimensions are positive. The absence of unitarity in these
models is due to the presence of anticommuting scalars and, at large $N$,
due to the violation of the unitarity bound for the scalar operator
$\sigma$.

We also considered the critical $\text{Sp}(N)$ models and saw that their
central charge indicates the existence of an interacting CFT in $d=7$.
Additionally, we saw that fixed points of \Lsym in the $1/N$ expansion in
$d=7$ have similar properties with the fixed points of $\Lag$ in $d=7$. One
cannot conclusively determine the relation of these fixed points at this
point, but it is tempting to suggest that they may be equivalent.  Further
support to this may come from a possible UV completion of these UV fixed
points starting from a theory in $d=8-\epsilon$.

It is important to investigate the way in which the theories proposed here
may play a role in the higher-spin~\cite{Fradkin:1987ks, Vasiliev:1990en,
Vasiliev:1992av, Vasiliev:2003ev} dS/CFT
correspondence~\cite{Strominger:2001pn, Anninos:2011ui}. These theories
fall in the class of models with weakly broken higher-spin symmetry, since
all higher-spin currents are nearly conserved in the large-$N$ limit.  In
$d=3$ this, along with the fact that in the AdS/CFT
correspondence~\cite{Maldacena:1997re, Gubser:1998bc, Witten:1998qj}
conserved vectors of the boundary theory correspond to gauge fields in the
bulk, led to the conjecture that the singlet sector of critical
$\text{O}(N)$ models is dual to interacting higher-spin theory in
$d=4$~\cite{Klebanov:2002ja}.  Since the Vasiliev equations are known for
all $d$~\cite{Vasiliev:2003ev} a similar result may apply for the $d=7$
models discussed here.

\newpage
\ack{I would like to thank the organizers of the Simons Summer Workshop
2015 where this work was initiated. I am grateful to Chris Beem, Clay
C\'{o}rdova, Jacques Distler, Thomas Dumitrescu, Igor Klebanov, David
Poland, Marco Serone, and Ran Yacoby for many helpful and stimulating
discussions. I also thank Simone Giombi, Igor Klebanov, and David Poland
for comments on the manuscript. My research is supported in part by the
National Science Foundation under Grant No.~1350180.}

\bibliography{symplectic_CFTs_6+epsilon}

\end{document}